\documentclass[11pt]{article}
\usepackage{mathrsfs}
\usepackage{amsmath}
\usepackage{amsfonts}
\usepackage{amssymb, amsmath, cite}
\usepackage{color}
\usepackage{graphicx}

\newtheorem{theorem}{Theorem}

\newcommand\be{\begin{equation}}
\newcommand\ee{\end{equation}}
\newcommand\ber{\begin{eqnarray}}
\newcommand\eer{\end{eqnarray}}
\newcommand\berr{\begin{eqnarray*}}
\newcommand\eerr{\end{eqnarray*}}
\newcommand\bea{\begin{eqnarray}}
\newcommand\eea{\end{eqnarray}}

\newcommand\dd{\mbox{d}}\newcommand\lm{\lambda}

\newcommand\pa{\partial}
\newcommand{\ud}{\mathrm{d}}
\newcommand{\nn}{\nonumber}

\title{Determination of Gap Solution and Critical\\ Temperature in Doped
Graphene Superconductivity}

\author{Chenmei Xu\\School of Mathematics and Statistics\\Henan University\\
Kaifeng, Henan 475004, PR China\\ \\
Yisong Yang\\Department of Mathematics\\Polytechnic School of Engineering\\ New York University\\Brooklyn, New York 11201, USA
\\  \&\\NYU-ECNU
Institute of Mathematical Sciences\\New York University - Shanghai\\3663 North Zhongshan Road, Shanghai 200062, PR China}

\date{}

\begin{document}
\maketitle
\begin{abstract}
 {It is shown that the gap solution and critical transition temperature are significantly enhanced by doping 
in a recently developed BCS formalism for graphene superconductivity in such a way that  positive gap and transition
temperature both occur in arbitrary pairing coupling as far as doping is present. The analytic construction of the BCS gap and transition temperature offers highly effective globally convergent iterative methods for the computation of these
quantities. A series of numerical examples are presented as illustrations consolidating the analytic understanding achieved.
 }
\medskip

\end{abstract}

\section{Introduction}
\setcounter{equation}{0}

Graphene research has been one of the most actively pursued topics in experimental and theoretical physics during the last decade since the work
of Geim and Novoselov \cite{GN}. Formed in a single-layer of carbon atoms bonded together in the shape of a hexagonal lattice, graphene is characterized by a list of salient physical properties such as high strength, high transparency,  high
heat conductivity, and almost zero electric resistance. Coupled with the abundancy of carbon in nature, graphene possesses
unlimited promises for future technology advancement and applications. Among the subjects of graphene research, graphene superconductivity has attracted considerable attention \cite{UCN,UN,TOB,WDLB,KS,G,KG,P,EE,NLC,KPH,HZ,UB,TK} from both experimental and theoretical physicists, encouraged by successes with other close relatives of graphene, such as graphite and fullerene. 
 More recently,  experimental breakthroughs unveiling the superconducting pairing/gap mechanism in calcium doped graphene are reported \cite{Fetal,Yetal}. Theoretically, such a superconducting mechanism was explored 
\cite{UCN,UN,KS,EE} much earlier, in view of the classical Bardeen--Cooper--Schrieffer (BCS) theory of phonon-based electron pairing \cite{BCS,AK,BFV,K}. In the context of graphene, however, electrons are Dirac fermions, meaning that they obey linear dispersion relations and move in extremely high velocities, inevitably leading to a wide range of new phenomena.
One of such new phenomena is that electron pairing would become unlikely in undoped (pure) graphene \cite{MN}
and superconductivity based on phonon and plasmon mediated pairing mechanisms requires doped (impure) samples \cite{WDLB,NLC,BS,B,BH}.

Naturally, doping-modified BCS equations \cite{UCN,KS} possess more complicated features than the classical BCS equation \cite{BCS}.
In particular, doping level would now make its appearance to influence other key physical quantities which characterize
superconductivity. Our main goal of the present work
 is to understand and describe the superconducting transition temperature
and  BCS gap as functions of the doping level, both analytically and numerically. Specifically, we will show that doping
enhances the transition temperature and BCS gap significantly so that they strictly increase with respect to the doping
level. In particular, we will show that, when there is no doping, a positive gap occurs only when the pairing coupling becomes sufficiently
strong, and that, when doping is present, a positive gap appears below the transition temperature no matter how weak
the pairing coupling is. Furthermore, as a by-product, we also obtain the behavior of the transition temperature and 
BCS gap with respect to the changes of other physical parameters such as the Fermi truncation energy and pairing coupling strength.

An outline of the rest of the paper is as follows. In \S 2 we introduce the doping-modified BCS equation and state our main
results which are divided into two theorems. In the first theorem, we recount a complete understanding of the 
doping-modified BCS equation regarding its gap solution and transition temperature versus various coupling parameters.
In the second theorem, we present globally convergent and monotonically iterative methods for the computation of the gap
solution and transition temperature of the doping-modified BCS equation. In \S 3 we establish the two theorems. In
\S 3.1 we begin by considering the zero-temperature equation.  Although the equation in this situation is rather simple
the study of it paves the path to the study of the full-setting finite-temperature equation in \S 3.2.
In particular we deduce the equation that determines the transition temperature by applying the fact that the BCS
gap vanishes at the transition temperature. In \S 3.3 we 
consider the transition-temperature equation. In \S 4 we present a series of numerical examples computed using the
globally convergent methods stated in \S 2 and developed in \S 3. These examples show the effectiveness of our methods and confirm
the analytic pictures described for the gap solution and transition temperature of the doping-modified BCS equation.
In \S 5 we draw conclusions.

\section{Doping-modified BCS equations, behavior of gap and transition temperature, and methods for computation}
\setcounter{equation}{0}

Below we shall consider the following physical parameters in the doping-modified BCS equations \cite{UCN,KS} in our study.
\begin{enumerate}
\item[] The absolute temperature $T\geq0$.

\item[] The gap function $\Delta\geq0$ which is known to be proportional to the order parameter describing 
the density of superconducting electron pairs in macroscopic theory so that $\Delta>0$ indicates the onset of
superconductivity and $\Delta=0$ implies the occurrence of normal conductivity.

\item[] The truncation energy $\xi_m>0$ which determines a
Fermi energy interval $|\xi_{\bf p}|<\xi_m$ in which attractive interaction
between electrons of opposite spins and momentum $\bf p$ is dominant.

\item[] The doping level  $\mu$ so that $\mu > 0$
indicates the electron doping and $\mu<0$ the hole doping.  As in \cite{UCN,KS}, we observe the low doping assumption 
\be\label{2.1}
|{\mu}|\leq \xi_{m}.
\ee

\item[] The pairing coupling constant $\lm>0$ measures the strength of attractive interaction within electron pairs.
\end{enumerate}

With the above notation, the finite-temperature doping-modified BCS gap equation derived in \cite{UCN,KS} reads
\be\label{2.2}
\frac{\xi_{m}}{\lambda}=2T\ln\frac{\cosh(\sqrt{\xi_{m}^{2}+\Delta^{2}}/2T)}{\cosh(\sqrt{\mu^{2}+\Delta^{2}}/2T)}
 +|{\mu}|\int_{0}^{|{\mu}|}\tanh\big(\frac{\sqrt{\xi^{2}+\Delta^{2}}}{2T}\big)\frac{\ud\xi}{\sqrt{\xi^{2}+\Delta^{2}}}.
\ee

Of course, in the zero-temperature limit, $T = 0$, the equation (\ref{2.2}) becomes \cite{UCN,KS},
\be\label{2.3}
\frac{\xi_{m}}{\lambda}=\sqrt{\xi_{m}^{2}+\Delta_{0}^{2}}-\sqrt{\mu^{2}+\Delta_{0}^{2}}
 +|{\mu}|\ln\frac{|{\mu}|+\sqrt{\mu^{2}+\Delta_{0}^{2}}}{\Delta_{0}},
\ee
after taking $T\to0$ in (\ref{2.2}).

Thus the doping-modified BCS equations are symmetric with respect to electron or hole doping.

At this moment, it will be instructive to recall the classical BCS equation \cite{BCS,AK,BFV,K}  in the same specific situation:
\be\label{2.4}
\Delta =V\int_{0}^{\xi_m}\tanh\left(\frac{\sqrt{\xi^{2}+\Delta^{2}}}{2T}\right)\frac{\Delta}{\sqrt{\xi^{2}+\Delta^{2}}} \,\ud\xi,
\ee
where $V>0$ is a constant, which allows us to see clearly the complication introduced by doping. Fortunately, we may 
extend the methods \cite{Y1,DY,Y2,Y3} for the classical BCS equation to achieve a completely understanding of the doping-modified BCS equations.

In fact, it is more convenient to recast (\ref{2.2}) into the `restored' BCS form (\ref{2.4}):
\be\label{2.5}
\frac{\xi_{m}}{\lambda}\Delta=2T\ln\frac{\cosh(\sqrt{\xi_{m}^{2}+\Delta^{2}}/2T)}{\cosh(\sqrt{\mu^{2}+\Delta^{2}}/2T)}\Delta
 +|{\mu}|\int_{0}^{|{\mu}|}\tanh\big(\frac{\sqrt{\xi^{2}+\Delta^{2}}}{2T}\big)\frac{\Delta}{\sqrt{\xi^{2}+\Delta^{2}}}\,\ud\xi.
\ee

For this equation, our results are summarized as follows.

\begin{theorem}\label{thm1} Consider the doping-modified finite-temperature BCS equation (\ref{2.5}) governing the gap function
$\Delta$ in terms of the pairing coupling constant $\lm>0$, the Fermi truncation energy $\xi_m>0$, the doping parameter $\mu$,
and the absolute temperature $T>0$.
\begin{enumerate}
\item[(i)] There exists a critical temperature $T_c>0$ such that (\ref{2.5}) possesses a unique positive solution,
say $\Delta=\Delta(\mu,T,\lm)$, for $T<T_c=T_c(\mu,\lm)$, and any $\mu\in(0,\xi_m]$ and
$\lm>0$, indicating the onset of superconductivity
with any pairing coupling strength in presence of doping. However, when $T\geq T_c$, the only nonnegative solution of (\ref{2.5}) is the trivial solution
$\Delta=0$, indicating the occurrence of normal conductivity at a sufficiently high
temperature. Besides, $\Delta$ and $T_c$ smoothly depend on
$\mu, T,\lm$.

\item[(ii)] Below $T_c$ there hold the monotonicity properties
\be
\frac{\pa\Delta}{\pa \mu}>0,\quad \frac{\pa\Delta}{\pa T}<0,\quad\frac{\pa\Delta}{\pa\lm}>0,\quad \mu>0,\quad T>0,
\quad\lm>0.
\ee

\item[(iii)] There holds the limiting property $\Delta\to0$ as $T\to T_c$.

\item[(iv)] The critical temperature $T_c>0$ is determined by the equation
\be\label{2.6}
\frac{\xi_{m}}{\lambda}=2T_{c}\ln\frac{\cosh{(\xi_{m}}/2T_{c})}{\cosh{(|\mu|/2T_{c})}}
+|{\mu}|\int_{0}^{|{\mu}|}\frac1{\xi}\tanh\left( \frac{\xi}{2T_{c}}\right)\, {\ud\xi}
\ee
uniquely and implicitly.

\item[(v)] The critical temperature $T_c$ given in (\ref{2.6}) enjoys the monotonicity properties
\be
\frac{\pa T_c}{\pa\mu}>0,\quad \frac{\pa T_c}{\pa\lm}>0,\quad \mu>0,\quad\lm>0.
\ee

\item[(vi)] The zero-temperature equation (\ref{2.3}) has a unique positive solution, $\Delta_0$, which is the limit
\be
\Delta_0=\Delta_0(\mu,\lm)\equiv\lim_{T\to0}\Delta(\mu,T,\lm),
\ee
and serves as the sharp upper bound for the finite-temperature gap $\Delta$: $\Delta_0>\Delta(\mu,T,\lm)\geq0$, $T>0$.
In addition, $\Delta_0$ smoothly and monotonically depend on $\mu\in (0,\xi_m]$ and $\lm>0$ so that
\be
\frac{\pa\Delta_0}{\pa\mu}>0,\quad \frac{\pa\Delta_0}{\pa\lm}>0.
\ee
Hence there hold the precise bounds
\be
\frac{\xi_m}2\left(\lm-\frac1\lm\right)=\Delta_0(0,\lm)<\Delta_0(\mu,\lm)<\Delta_0(\xi_m,\lm)=\frac{\xi_m}{\sinh\left(\frac1\lm\right)},\quad \mu\in(0,\xi_m),\quad \lm>0,
\ee
which implies in particular that a positive gap at zero doping requires a strong pairing coupling, $\lm>1$, although in the finite
doping $\mu>0$, however, a positive gap occurs at any finite pairing coupling, $\lm>0$.

\item[(vii)] At zero doping $\mu=0$ and finite-temperature $T>0$,
a positive gap can only happen when $\lm>1$, as in the zero-temperature case.
\end{enumerate}
\end{theorem}

For the original doping-modified equation (\ref{2.2}), we see that there is no solution for $T>T_c$ and that the only solution
is the zero solution for $T=T_c$, as a consequence of (\ref{2.6}).

The gap solution and critical temperature described above can all be obtained using globally convergent iterative methods
as presented below.

\begin{theorem}\label{thm2}
The gap solutions and critical temperature stated in Theorem \ref{thm1} for $\mu>0$ may all be obtained iteratively 
and monotonically by globally
convergent methods below.
\begin{enumerate}
\item[(i)] For the zero-temperature equation (\ref{2.3}), define 
\be\label{2.12}
u_{n+1}=\frac\lm{\xi_m}u_n\left(\sqrt{\xi_m^2+u^2_n}-\sqrt{\mu^2+u^2_n}+\mu\ln\frac{\mu+\sqrt{\mu^2+u_n^2}}{u_n}\right),\quad n=0,1,2,\dots.
\ee
\item[(ii)] For the finite-temperature equation (\ref{2.2}), define
\bea\label{2.13}
u_{n+1}&=&\frac\lm{\xi_m}u_n\left(
2T\ln\frac{\cosh(\sqrt{\xi_{m}^{2}+u_n^{2}}/2T)}{\cosh(\sqrt{\mu^{2}+u_n^{2}}/2T)}
 +{\mu}\int_{0}^{{\mu}}\tanh\left(\frac{\sqrt{\xi^{2}+u_n^{2}}}{2T}\right)\frac{\ud\xi}{\sqrt{\xi^{2}+u_n^{2}}}\right),\nn\\
 n&=&0,1,2,\dots.
\eea
\item[(iii)] For the critical temperature equation (\ref{2.6}), define
\be\label{2.14}
\tau_{n+1}=\frac\lm{\xi_m}\tau_n\left(
2\tau_n\ln\frac{\cosh{(\xi_{m}}/2\tau_n)}{\cosh{(\mu/2\tau_n)}}
+{\mu}\int_{0}^{{\mu}}\frac1{\xi}\tanh\left( \frac{\xi}{2\tau_n}\right)\, {\ud\xi}\right),\quad n=0,1,2,\dots.
\ee
\end{enumerate}
For any initial state $u_0>0$, the sequence $\{u_n\}$
defined by (\ref{2.12}) converges monotonically to a positive limit $u_*$ which is the 
unique gap solution of the zero-temperature BCS equation (\ref{2.3}); if $0<T<T_c$,
the sequence $\{u_n\}$
defined by (\ref{2.13}) converges monotonically to a positive limit $u_*$ which is the 
unique gap solution of the finite-temperature BCS equation (\ref{2.2}); however, if $T\geq T_c$, the limit
vanishes, $u_*=0$,
which indicates that (\ref{2.2}) has no solution. For any $\tau_0>0$, the sequence $\{\tau_n\}$ defined by (\ref{2.14})
converges monotonically to the  critical transition temperature $T_c>0$ of the finite-temperature BCS equation
which is the unique positive solution of the critical-temperature equation (\ref{2.6}). In all these cases, the convergence possesses a linear rate.
\end{theorem}

These theorems will be established in the next section.

\section{Construction of gap solutions and critical temperature}
\setcounter{equation}{0}

In this section we establish Theorems \ref{thm1} and \ref{thm2}. We first solve 
the  zero-temperature equation (\ref{2.3}) in the beginning subsection.
Although this equation is simple, the insight gained will allow us to tackle the full finite-temperature equation (\ref{2.2})
and the critical-temperature equation (\ref{2.6}) in the subsequent subsections.

\subsection{The zero-temperature equation}

We only need to consider the case when $\mu>0$. Thus, rewrite (\ref{2.3}) as
\be\label{3.1}
\frac{\xi_{m}}{\lambda}=f(\mu,\Delta_{0})\equiv\sqrt{\xi_{m}^{2}+\Delta_{0}^{2}}-\sqrt{\mu^{2}+\Delta_{0}^{2}}
 +{\mu}\ln\left(\frac{{\mu}+\sqrt{\mu^{2}+\Delta_{0}^{2}}}{\Delta_{0}}\right).
\ee
Then we have
\bea
\frac{\pa f}{\pa\Delta_0}&=&\Delta_0\left(\frac1{\sqrt{\xi^2_m+\Delta_0^2}}-\frac1{\sqrt{\mu^2+\Delta_0^2}}\right)-
\frac{\mu^2}{\Delta_0\sqrt{\mu^2+\Delta_0^2}},\label{3.2}\\
\frac{\pa f}{\pa\mu}&=&\ln\left(\frac{{\mu}+\sqrt{\mu^{2}+\Delta_{0}^{2}}}{\Delta_{0}}\right).\label{3.3}
\eea
Hence, applying the weak doping condition (\ref{2.1})
in (\ref{3.2}), we get $\frac{\pa f}{\pa\Delta_0}<0$. Moreover, we also know that
\be\label{3.4}
\lim_{\Delta_0\to0}f(\mu,\Delta_0)=\infty,\quad
\lim_{\Delta_0\to\infty}f(\mu,\Delta_0)=0.
\ee
So (\ref{3.1}) has a unique positive solution which may be obtained by all kinds of standard means and is known to be
a smooth function of the variables $\mu,\lm,\xi_m$. Since for $\mu>0$ we have
$
\frac{\pa f}{\pa\mu}>0
$ by (\ref{3.3}), we may apply the implicit function theorem to obtain the property
\be
\frac{\dd\Delta_0}{\dd\mu}>0.
\ee
That is, a high level of doping results in a larger gap. Moreover, when $\mu=0$, (\ref{3.1}) gives us
\be
\Delta_0=\frac{\xi_m}2\left(\lm-\frac1\lm\right),
\ee
which is consistent with the condition $\Delta_0\geq0$ when $\lm\geq1$ (cf. \cite{KS}). When $\mu=\xi_m$, (\ref{3.1}) results in
\be
\Delta_0=\frac{\xi_m}{\sinh(\frac1\lm)}.
\ee
Consequently, we see that the monotonicity leads to the explicit range for the gap $\Delta_0$ as follows:
\be\label{3.8x}
\frac{\xi_m}2\left(\lm-\frac1\lm\right)\leq\Delta_0\leq \frac{\xi_m}{\sinh(\frac1\lm)},\quad 0\leq\mu\leq\xi_m.
\ee

(It should be noted that the self-consistency condition
\be
\frac{1}2\left(\lm-\frac1\lm\right)<\frac{1}{\sinh(\frac1\lm)},\quad \lm>0,
\ee
implied by (\ref{3.8x}) is always valid unconditionally for any $\lm$ as may be checked directly, although it is not so obvious.)

We now aim to construct the unique solution of (\ref{3.1}) for $\mu\in (0,\xi_m)$. The property (\ref{3.4})
however prevents a direct iterative scheme to be carried out and some suitable modification of the equation is needed. For this purpose, we use $u$ to denote
$\Delta_0$ and rewrite (\ref{3.1}) as a fixed-point equation in the form
\begin{eqnarray}
u=F(u)\equiv \frac{\lambda}{\xi_{m}}uf(u),\quad u>0,\label{3.9}
\end{eqnarray}
where we have suppressed the dependence of $f$ on the parameter $\mu$ for simplicity of notation.

For our method to work, it is crucial to note that the function $g(u)=uf(u)$ is monotone increasing in $u > 0$. To see this
not-so-obvious but elementary property, we compute to get
\bea
g'(u)&=&\frac{\xi_m^2+2u^{2}}{\sqrt{\xi_{m}^{2}+u^2}}-2\sqrt{\mu^{2}+u^{2}}
 +{\mu}\ln\left(\frac{{\mu}+\sqrt{\mu^{2}+u^{2}}}{u}\right),\\
g''(u)&=&\frac{3u\xi_{m}^{2}+2u^{3}}{(\xi_{m}^{2}+u^{2})^{\frac{3}{2}}}-\frac{\mu^{2}+2u^{2}}{u\sqrt{\mu^{2}+u^{2}}},\\
g'''(u)&=&\frac{3\xi_{m}^{4}}{(\xi_{m}^{2}+u^{2})^{\frac{5}{2}}}+\frac{\mu^{4}}{u^{2}(\mu^{2}+u^{2})^{\frac{3}{2}}}.
\eea
From $
g'''(u)>0,\lim_{u \rightarrow 0}g''(u)=-\infty, \lim_{u\rightarrow \infty }g''(u)=0$
we have $
g''(u)<0$. Thus, from $
\lim_{u \rightarrow 0}g'(u)=\infty, \lim_{u\rightarrow \infty }g'(u)=0$,
we get
$g'(u)>0$ ($u>0$), as claimed.

On the other hand, in view of the property (\ref{3.4}), we see that
\begin{eqnarray}
\lim_{u \rightarrow 0}\frac{F(u)}{u}=\infty,\quad \lim_{u\rightarrow \infty }\frac{F(u)}{u}=0.
\end{eqnarray}
Hence there are numbers $\varepsilon_0 > 0 $
and $\delta_0 > 0$  with $ \varepsilon < \delta$ such that
\begin{eqnarray}
\varepsilon < F(\varepsilon),\quad \varepsilon\in (0,\varepsilon_0); \quad F(\delta)<\delta,\quad \delta\in (\delta_0,\infty).
\end{eqnarray}
In other words, the equation $u=F(u)$ ($u>0$) has sufficiently small subsolutions and sufficiently large supersolutions. Therefore we can conclude that
 the iterative sequence
\be\label{3.15}
u_{n+1}=F(u_n),\quad n=0,1,2,\dots, 
\ee
converges to the unique solution, say $u_*$, of the equation $u=F(u)$ ($u>0$) for any choice of the initial state $u_0\in (0,\infty)$.
Moreover, the sequence enjoys the property that
\bea
&&u_n<u_{n+1},\quad n=0,1,2,\dots,\quad \mbox{if }u_0<F(u_0); \\
&& u_n>u_{n+1},\quad n=0,1,2,\dots,\quad \mbox{if }u_0>F(u_0).
\eea
In either case, using the relation (\ref{3.15}),  we obtain
\begin{eqnarray}
\lim_{n\rightarrow \infty}\frac{u_{n+1}-u_{*}}{u_{n}-u_{*}}=\frac{\lm}{\xi_m}\lim_{n\to\infty}
\left(\frac{g(u_n)-g(u_*)}{u_n-u_*}\right)=\frac{\lambda}{\xi_{m}}g'(u_{*}). 
\end{eqnarray}
Since $g'(u_{*})>0$, we see that $\{u_n\}$ converges to $u_*$ with a linear rate.

\subsection{The finite-temperature equation}

We now conduct a study of the full equation (\ref{2.2}) with $\mu\geq0$. For convenience, we set
 \be
f(\mu,\Delta, T)=2T\ln\frac{\cosh(\sqrt{\xi_{m}^{2}+\Delta^{2}}/2T)}{\cosh(\sqrt{\mu^{2}+\Delta^{2}}/2T)}
 +{\mu}\int_{0}^{{\mu}}\tanh\frac{\sqrt{\xi^{2}+\Delta^{2}}}{2T}\frac{\dd\xi}{\sqrt{\xi^{2}+\Delta^{2}}}.
\ee
Then it can be checked that
\begin{eqnarray}
\lim_{T \rightarrow \infty }f(\mu,\Delta,T)&=&0,\label{3.20}\\
\lim_{\Delta \rightarrow 0} f(\mu,\Delta,T)&=&2T\ln\frac{\cosh{(\xi_{m}}/2T)}{\cosh{(\mu/2T)}}
+{\mu}\int_{0}^{{\mu}}\tanh \frac{\xi}{2T}\frac{\dd\xi}{\xi}, \quad T>0,\label{3.21}\\
\lim_{\Delta \rightarrow \infty }f(\mu,\Delta,T)&=&0, \quad T>0.\label{3.22}
\end{eqnarray}
Besides, we also have
\bea
\frac{\pa f}{\pa\mu}&=&\int_0^\mu \tanh\frac{\sqrt{\xi^{2}+\Delta^{2}}}{2T}\frac{\dd\xi}{\sqrt{\xi^{2}+\Delta^{2}}},\\
 \frac{\partial f}{\partial \Delta}&=&\Delta\left(\frac1{\sqrt{\xi_m^2+\Delta^2}}\tanh\frac{\sqrt{\xi_m^2+\Delta^2}}{2T}
-\frac1{\sqrt{\mu^2+\Delta^2}}\tanh\frac{\sqrt{\mu^2+\Delta^2}}{2T}\right)\nn\\
&&+\frac\mu2\int_0^\mu\frac\Delta{(\xi^2+\Delta^2)^{\frac32}\cosh^2\frac{\sqrt{\xi^2+\Delta^2}}{2T}}
\left(\frac{\sqrt{\xi^2+\Delta^2}}T-\sinh \frac{\sqrt{\xi^2+\Delta^2}}T\right)\,\dd\xi.\label{3.25}
\eea
Thus $\frac{\pa f}{\pa\mu}>0$ and $\frac{\pa f}{\pa\Delta}<0$. In addition, we have
\be
\frac{\partial f}{\partial T} =q(T)-\frac{\mu}{2T^2}\int_0^\mu\frac1{\cosh^2\frac{\sqrt{\xi^2+\Delta^2}}{2T}}\,\dd\xi,
\ee
where
\bea
q(T)&=&2\ln\frac{\cosh(\sqrt{\xi_{m}^{2}+\Delta^{2}}/2T)}{\cosh(\sqrt{\mu^{2}+\Delta^{2}}/2T)}\nn\\
&&
-\frac1T\left(\sqrt{\xi_m^2+\Delta^2}\tanh\frac{\sqrt{\xi_{m}^{2}+\Delta^{2}}}{2T}
-\sqrt{\mu^2+\Delta^2}\tanh\frac{\sqrt{\mu^{2}+\Delta^{2}}}{2T}\right)\nn\\
&=&\frac1T\left(\sqrt{\xi_m^2+\Delta^2}-\sqrt{\mu^2+\Delta^2}\right)\tanh\eta\nn\\
&&-\frac1T\left(
\sqrt{\xi_m^2+\Delta^2}\tanh\frac{\sqrt{\xi_m^2+\Delta^2}}{2T}-\sqrt{\mu^2+\Delta^2}\tanh\frac{\sqrt{\mu^2+\Delta^2}}{2T}\right)\nn\\
&=&\frac{\sqrt{\xi_m^2+\Delta^2}}T\left(\tanh\eta-\tanh\frac{\sqrt{\xi_m^2+\Delta^2}}{2T}\right)\nn\\
&&+\frac{\sqrt{\mu^2+\Delta^2}}T\left(\tanh\frac{\sqrt{\mu^2+\Delta^2}}{2T}-\tanh \eta\right),
\eea
with $\eta\in\left(\frac{\sqrt{\mu^2+\Delta^2}}{2T},\frac{\sqrt{\xi_m^2+\Delta^2}}{2T}\right)$, which indicates  $q(T)<0$. So we get $\frac{\pa f}{\pa T}<0$.

Consequently, using the implicit function theorem, we see that the gap solution, say $\Delta$, to the BCS equation
\be\label{3.27}
\frac{\xi_m}\lm =f(\mu,\Delta,T)
\ee
is unique (if it exists) and depends on the doping parameter $\mu$ and absolute temperature $T$ smoothly and
monotonically:
\be
\frac{\pa\Delta}{\pa\mu}>0,\quad\frac{\pa \Delta}{\pa T}<0.
\ee
In particular, as in the zero temperature situation, doping enhances the BCS gap.

Therefore, for fixed finite temperature $T>0$, the smallest gap occurs at the zero doping, $\mu=0$. Thus, from (\ref{3.27}), we see
that $\Delta|_{\mu=0}=\Delta(0,T)$ is determined by the equation
\be\label{3.30}
\frac{\cosh\frac{\sqrt{\xi_m^2+\Delta^2}}{2T}}{\cosh\frac\Delta{2T}}=\exp\left({\frac{\xi_m}{2\lm T}}\right).
\ee

It should be noted that,
as in the zero-temperature situation, (\ref{3.30}) may not always possess a solution. To see this fact, 
denote by $h(\Delta)$ the left-hand side of (\ref{3.30}) and note that $h(\Delta)$
monotonically decreases in $\Delta\geq0$.  Since $h(0)=\cosh\left(\frac{\xi_m}{2T}\right)$ and $ h(\infty)=1$, we see
that (\ref{3.30}) has a solution $\Delta>0$ if and only if
\be\label{3.31}
\exp\left({\frac{\xi_m}{2\lm T}}\right)<\cosh\left(\frac{\xi_m}{2T}\right).
\ee
Thus, in order to ensure (\ref{3.31}), we need to impose the strong
pairing coupling condition $\lm>1$, as in the zero-temperature equation case. See also \cite{KS}.

The largest gap on the other hand occurs at the maximum doping, $\mu=\xi_m$, which is 
$\Delta|_{\mu=\xi_m}=\Delta(\xi_m,T)$ and determined by the
classical BCS equation \cite{BCS}:
\be
\frac1\lm=\int_{0}^{{\xi_m}}\tanh\frac{\sqrt{\xi^{2}+\Delta^{2}}}{2T}\frac{\dd\xi}{\sqrt{\xi^{2}+\Delta^{2}}}.
\ee

Now for fixed doping parameter $\mu$, let $\Delta(\mu,T)$ solve (\ref{3.27}). Then we have the uniform bounds
\be
\Delta(0,T)\leq\Delta(\mu,T)\leq \Delta(\xi_m,T),\quad 0\leq\mu\leq\xi_m,\quad T>0.
\ee

From (\ref{3.20})--(\ref{3.22}) and the above results, we arrive at the following conclusions.

\begin{enumerate}
\item[(i)] The finite-temperature BCS equation (\ref{3.27}) has a positive gap solution $\Delta$ if and only if $\mu$ and $T$
satisfy
\be\label{3.32}
\frac{\xi_m}\lm<2T\ln\frac{\cosh{(\xi_{m}}/2T)}{\cosh{(\mu/2T)}}
+{\mu}\int_{0}^{{\mu}}\tanh \frac{\xi}{2T}\frac{\dd\xi}{\xi}.
\ee
(Note that this condition contains (\ref{3.31}).)
Therefore, for fixed $\mu\in [0,\xi_m]$, there is critical temperature $T_c(\mu)$, such that (\ref{3.27}) has a positive
solution $\Delta$ when $T<T_c(\mu)$ and that (\ref{3.27}) has no solution when $T>T_c(\mu)$.

\item[(ii)] If a positive gap solution of (\ref{3.27}) exists, it must be unique.

\item[(iii)] For $T<T_c(\mu)$ the gap solution $\Delta(\mu,T)$ is monotone decreasing with respect to $T$ and
\be
\lim_{T\to (T_c(\mu))^-}\Delta(\mu,T)=0.\label{3.33}
\ee
\end{enumerate}

We need only to establish the limit (\ref{3.33}). Assume otherwise
\be
\lim_{T\to (T_c(\mu))^-}\Delta(\mu,T)=\Delta_c(\mu)>0.\label{3.34}
\ee
Then (\ref{3.27}) leads to
\bea
\frac{\xi_m}\lm&=&f(\mu,\Delta_c(\mu),T_c(\mu))<f(\mu,0,T_c(\mu))\nn\\
&=&2T_c(\mu)\ln\frac{\cosh{(\xi_{m}}/2T_c(\mu))}{\cosh{(\mu/2T_c(\mu))}}
+{\mu}\int_{0}^{{\mu}}\tanh \frac{\xi}{2T_c(\mu)}\frac{\dd\xi}{\xi}.
\eea
By continuity we can choose $T>T_c(\mu)$ to achieve (\ref{3.32}) which conflicts with the definition of $T_c(\mu)$
described.

Below we proceed to obtain a globally convergent iterative computational scheme for the gap solution of the
finite-temperature BCS equation
as in the zero-temperature situation.

As in \S 3.1, use $u$ to denote $\Delta$ and suppress the fixed parameters $\mu$ and $T$ in the function $f$:
\be
f(u)=2T\ln\frac{\cosh(\sqrt{\xi_{m}^{2}+u^{2}}/2T)}{\cosh(\sqrt{\mu^{2}+u^{2}}/2T)}
 +{\mu}\int_{0}^{{\mu}}\tanh\frac{\sqrt{\xi^{2}+u^{2}}}{2T}\frac{\dd\xi}{\sqrt{\xi^{2}+u^{2}}}.
\ee
Hence (\ref{3.27}) assumes the form (\ref{3.9}) as before. For convenience, set
\be
g(u)=u\ln\frac{\cosh(\sqrt{\xi_{m}^{2}+u^{2}}/2T)}{\cosh(\sqrt{\mu^{2}+u^{2}}/2T)}.
\ee
Then
\bea
2Tg'(u)&=&2T\ln\frac{\cosh(\sqrt{\xi_{m}^{2}+u^{2}}/2T)}{\cosh(\sqrt{\mu^{2}+u^{2}}/2T)}\nn\\
&&+{u^2}\left(\frac1{\sqrt{\xi_m^2+u^2}}\tanh\frac{\sqrt{\xi_m^2+u^2}}{2T}
-\frac1{\sqrt{\mu^2+u^2}}\tanh\frac{\sqrt{\mu^2+u^2}}{2T}\right)\nn\\
&=&\left(\sqrt{\xi_m^2+u^2}-\sqrt{\mu^2+u^2}\right)\tanh \eta\nn\\
&&+{u^2}\left(\frac1{\sqrt{\xi_m^2+u^2}}\tanh\frac{\sqrt{\xi_m^2+u^2}}{2T}
-\frac1{\sqrt{\mu^2+u^2}}\tanh\frac{\sqrt{\mu^2+u^2}}{2T}\right)\nn\\
&>&\left(\sqrt{\xi_m^2+u^2}-\sqrt{\mu^2+u^2}\right)\tanh\frac{\sqrt{\mu^2+u^2}}{2T} \nn\\
&&+{u^2}\left(\frac1{\sqrt{\xi_m^2+u^2}}\tanh\frac{\sqrt{\xi_m^2+u^2}}{2T}
-\frac1{\sqrt{\mu^2+u^2}}\tanh\frac{\sqrt{\mu^2+u^2}}{2T}\right)\nn\\
&>&\left(\frac{(\xi_m^2+2u^2)}{\sqrt{\xi_m^2+u^2}}-\frac{(\mu^2+2u^2)}{\sqrt{\mu^2+u^2}}\right)\tanh\frac{\sqrt{\mu^2+u^2}}{2T}>0,
\eea
where $\eta\in\left(\frac{\sqrt{\mu^2+u^2}}{2T},\frac{\sqrt{\xi_m^2+u^2}}{2T}\right)$.
Thus $uf(u)$ ($u>0$) strictly increases as in the zero-temperature situation. Therefore we conclude as before that the iterative
sequence defined by the scheme
\be\label{3.39}
u_{n+1}=\frac\lm{\xi_m} u_n f(u_n),\quad n=0,1,2,\dots, 
\ee
is monotonically convergent for any initial state $u_0>0$ and whether it increases or decreases depends on whether
$u_1>u_0$ or $u_1<u_0$. Moreover, set
\be 
u_*=\lim_{n\to\infty} u_n.
\ee
Then it is the unique positive gap solution of (\ref{3.27}) when $0<T<T_c(\mu)$ and $u_*=0$ when $T\geq T_c(\mu)$.
In other words, the positivity of $u_*$ is indicative of whether or not $T$ is below $T_c(\mu)$.

To end this subsection, we rewrite (\ref{3.27}) as
\be
g(\mu,\Delta,T,\lm)\equiv f(\mu,\Delta, T)-\frac{\xi_m}\lm.
\ee
Since $\frac{\pa g}{\pa\lm}=\frac{\xi_m}{\lm^2}>0$, we see that the positive gap solution of (\ref{3.27}) enjoys the property
\be
\frac{\pa\Delta}{\pa\lm}>0.
\ee
Thus a great pairing strength results in a larger BCS gap. This is an anticipated phenomenon in superconductivity theory.
In the next subsection, we will see that the same property is shared by the transition temperature as well.

\subsection{The critical temperature}

Use $T_c=T_c(\mu)>0$ and let $T\to T_c$ in (\ref{3.27}). Applying the fact $\Delta(\mu,T)\to 0$ as $T\to T_c$ obtained
in \S 3.2, we arrive at the following equation
\be\label{3.41}
\frac{\xi_m}\lm=f(\mu,T_c)\equiv 2T_c\ln\frac{\cosh{(\xi_{m}}/2T_c)}{\cosh{(\mu/2T_c)}}
+{\mu}\int_{0}^{{\mu}}\tanh\left( \frac{\xi}{2T_c}\right)\frac{\dd\xi}{\xi},
\ee
for the determination of $T_c$. Furthermore, we have
\bea
\frac{\pa f}{\pa\tau}(\mu,\tau)&=&2\ln\frac{\cosh\left(\frac{\xi_m}{2\tau}\right)}{\cosh\left(\frac{\mu}{2\tau}\right)}-\frac{\xi_m}\tau\tanh\left(\frac{\xi_m}{2\tau}\right)+\frac\mu\tau\tanh\left(\frac\mu{2\tau}\right)-\frac\mu{2\tau^2}\int_0^\mu
\frac{\dd\xi}{\cosh^2\left(\frac\xi{2\tau}\right)}\nn\\
&=&\frac1\tau(\xi_m-\mu)\tanh\eta-\frac{\xi_m}\tau\tanh\left(\frac{\xi_m}{2\tau}\right)+\frac\mu\tau\tanh\left(\frac\mu{2\tau}\right)-\frac\mu{2\tau^2}\int_0^\mu
\frac{\dd\xi}{\cosh^2\left(\frac\xi{2\tau}\right)}\nn\\
&=&\frac{\xi_m}\tau\left(\tanh\eta-\tanh\frac{\xi_m}{2\tau}\right)+\frac\mu\tau\left(\tanh\frac\mu{2\tau}-\tanh\eta\right)
-\frac\mu{2\tau^2}\int_0^\mu
\frac{\dd\xi}{\cosh^2\left(\frac\xi{2\tau}\right)}\nn\\
&<&0,
\eea
where $\eta\in\left(\frac\mu{2\tau},\frac{\xi_m}{2\tau}\right)$. Note also that
\be\label{3.43}
\lim_{\tau\to0}f(\mu,\tau)=\infty,\quad\lim_{\tau\to\infty}f(\mu,\tau)=0.
\ee
Hence the existence and uniqueness of the critical temperature $T_c>0$ for any given $\mu$ is ensured.
Besides, there holds
\be
\frac{\pa f}{\pa\mu}=\int_0^\mu\tanh\left(\frac\xi{2\tau}\right)\frac{\dd\xi}\xi>0.
\ee
Hence in view of the implicit function theorem we see that $T_c$ as a smooth function of the doping parameter $\mu$ enjoys the property
\be
\frac{\dd T_c}{\dd\mu}>0.
\ee
In other words, doping, whether electron or hole doping, enhances the critical temperature, which is analoguous to the fact that doping enhances the BCS gap observed earlier.

Similarly, we have
\be
\frac{\dd T_c}{\dd\lm}>0.
\ee
This result is natural which simply says a stronger pairing mechanism gives rise to a higher transition temperature.

To compute the critical temperature $T_c$, we rewrite (\ref{3.41}) in the form of a fixed-point equation:
\be\label{3.46}
\tau=F(\tau)\equiv \frac\lm{\xi_m}\tau f(\tau),\quad \tau>0,
\ee
where we have suppressed the parameter $\mu$ for convenience (e.g., $f(\tau)=f(\mu,\tau)$).

The property (\ref{3.43}) indicates that the equation (\ref{3.46}) has sufficiently small subsolutions and sufficiently large
supersolutions. Hence we may invoke the iterative sequence as before
\be\label{3.47}
\tau_{n+1}=F(\tau_n),\quad \tau_0>0,
\ee
whose limit solves the fixed-point equation (\ref{3.46}) and is the crirical transition temperature of the BCS gap equation:
\be
T_c=\lim_{n\to\infty} \tau_n.
\ee

In the actual implementation of the scheme (\ref{3.47}), we may first use some standard methods (e.g., the bisection method
applied on $f(\tau)$)
to approximate the fixed-point of $F(\tau)$ and then start the iteration, to accelerate the convergence.

\section{Numerical examples}
\setcounter{equation}{0}\setcounter{figure}{0}

In this section we present some numerical examples which serve two purposes. The first is to illustrate the efficiency of our iterative methods 
established in the previous sections. The second is to demonstrate and confirm the detailed anticipated behavior of the BCS gap and
critical temperature with respect to the changes of various physical parameters in the theory including
the absolute temperature $T$, the doping level $\mu$, the Fermi truncation energy $\xi_m$, and the
pairing coupling strength $\lm$.

\subsection{Zero-temperature gap solutions}

To begin our illustration, we take $\xi_{m}=2, \lambda=0.8$, and $ \mu=1$ in the zero-temperature BCS equation
(\ref{2.3}) and consider the iterative scheme (\ref{2.12}). 
The numerical accuracy threshold of our approximation sequence is set to be
\be\label{4.1}
|u_{n}-u_{n-1}|<10^{-8}.
\ee
In other words, we terminate the iteration and accept $ u_{n} $ to be an approximate solution of the
BCS equation (\ref{2.3}) when (\ref{4.1}) is attained.

\begin{enumerate}

\item[(i)] First we take $u_0=2.5$. The computation terminates at $n=34$ and renders
\be\label{4.2}
u_{34}=0.4465105266837066.
\ee

\item[(ii)] Next we take $u_0=0.05$. The computation terminates at $n=37$ and renders
\be\label{4.3}
u_{37}= 0.4465105023252628.
\ee
\end{enumerate}

\begin{figure}[htb]
   \centering
   \includegraphics[scale=0.8]{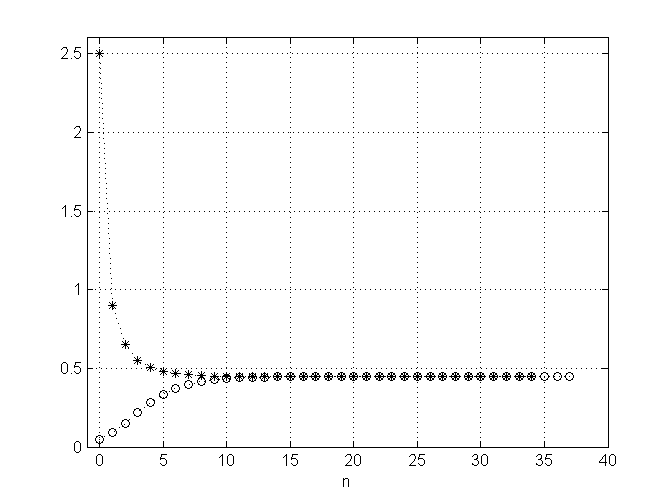}
   \caption{The plots of the iterative sequences initiating at $u_0=2.5$ and $u_0=0.05$, respectively. Although the initial states are far apart, the sequences converge quickly to yield the solution.}\label{F1}
\end{figure}

The monotonic behavior of the iterative sequences is shown in Figure \ref{F1}.

It is seen that the two sequences quickly level after 10 iterations. Moreover, the results (\ref{4.2}) and (\ref{4.3})
suggest that the gap solution is given by
\be
\Delta_0=u_*=0.4465105
\ee
within 7 digits after the decimal point.

\subsection{Finite-temperature gap solutions}

We take $T=0.9$,  $\xi_{m}=2, \lambda=2,$ and $\mu=1$ in the finite-temperature BCS equation (\ref{2.2}) and apply the
scheme (\ref{2.13}). The integral in the equation here, as well as in the sequel, is computed using
the trapezoidal rule  with 100 equidistant subdivisions for the integration interval. We again observe the termination threshold
(\ref{4.1}).

\begin{enumerate}
\item[(i)]  First we take $u_0=2.5$. The computation terminates at $n=46$  and yields
\be\label{4.4}
u_{46}=1.606160448787221.
\ee

\item[(ii)] We then take $u_0=1.2$. 
The computation terminates at $ n=48$  and yields
\be\label{4.5}
u_{48}= 1.606160441996964.
\ee
\end{enumerate}

\begin{figure}[htbp]
       \setlength{\abovecaptionskip}{0pt}
       \setlength{\belowcaptionskip}{0pt}
   \centering
   \includegraphics[scale=0.8]{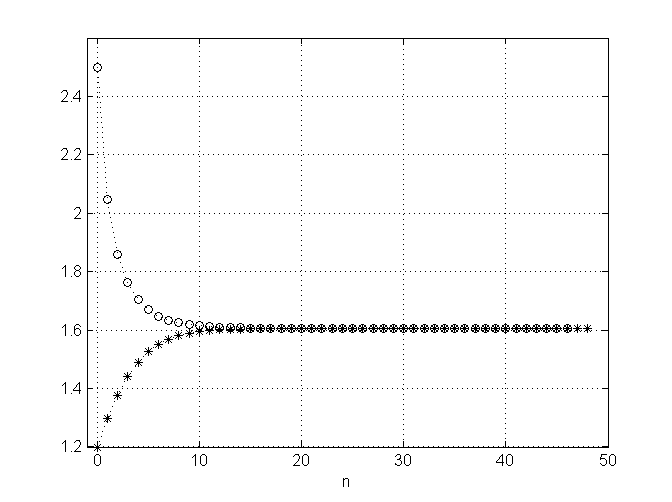}
   \caption{The plots of the iterative sequences initiating at $u_0=2.5$ and $u_0=1.2$, respectively. The initial states are again taken to be far apart but the generated sequences converge quickly to the solution, showing that our method is insensitive to the choice of initial state and rather robust.}\label{F2}
\end{figure}

The monotonically convergent behavior of the two sequences is shown in Figure \ref{F2}.

We observe that the two sequences level after 30 iterations. Furthermore, the results (\ref{4.4}) and (\ref{4.5}) suggest
that, within 6 digits after the decimal point, the gap solution is
\be
\Delta =u_*=1.606160.
\ee

\subsection{Computation of critical temperature}

Take $\xi_{m}=2, \lambda=2, \mu=1$ and use the iterative sequence defined by (\ref{2.14}) to approximate the solution $T_{c}$
of the critical-temperature equation (\ref{2.6}). 
As in the computation of the gap solutions, we set the termination threshold to be
\be\label{4.8}
|\tau_{n}-\tau_{n-1}|<10^{-8}.
\ee

\begin{enumerate}
\item[(i)] For $\tau_0=5$ the iteration terminates at
 $n=11$ and gives us the result
\be\label{4.9}
\tau_{12}=  1.140654043939101.
\ee

\item[(ii)] For $\tau_0=0.5$ the iteration terminates at $n=12$ and gives us the result
\be\label{4.10}
\tau_{13}=1.140654042231516.
\ee
\end{enumerate}

\begin{figure}[htbp]
   \centering
   \includegraphics[scale=0.8]{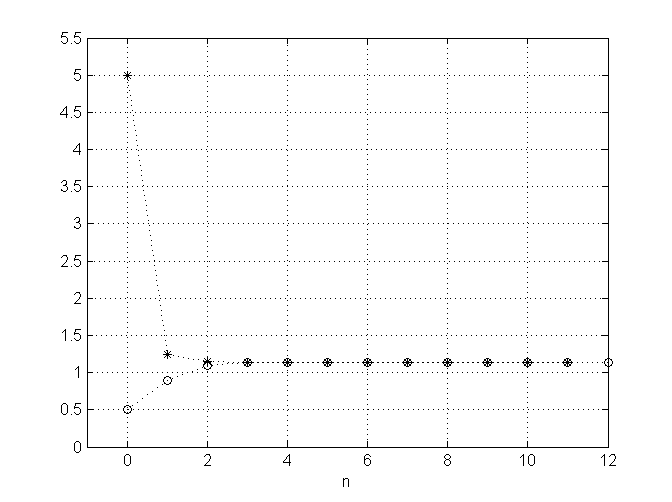}
   \caption{The plots of the iterative sequences initiating at $\tau_0=5$ and $\tau_0=0.5$, respectively. The effectiveness of our method for the determination of the critical temperature is demonstrated by the fast global convergence of the two
iterative sequences starting from initial states which are purposely chosen far apart.}\label{F3}
\end{figure}

The behavior of the iterative sequences starting from $\tau_0=5$ and $\tau_0=0.5$ are
exhibited in Figure \ref{F3}. 

It is clearly seen that the two sequences level very fast after just a few iterations, although they start from
the initial values which are chosen rather far apart. The results (\ref{4.9}) and (\ref{4.10}) indicate that the critical
transition temperature $T_c$ may be determined with 8 digits after the decimal point to be
\be\label{4.11}
T_c=1.14065404.
\ee

\subsection{Dependence of gap and critical temperature on various parameters}

In this subsection, we present a series of numerical results confirming the dependence of the BCS gap and the critical
transition temperature on the physical parameters $\xi_m,\mu,\lm$, obtained using the iterative methods established.

 We first consider the critical temperature problem.   The iterative scheme will render $\tau_n$ as
the solution $T_c$ when it meets the termination criterion
\be
|\tau_n-\tau_{n-1}|<10^{-10}.
\ee
We will only display 4 digits after the decimal point for all results.

\begin{enumerate}

\item[(i)] Take $\xi_m=2$ and $\mu=1$. Figure \ref{F4} shows the monotone dependence of $T_c$ on $\lm$. When
$\lm=1.1$, the scheme terminates after 28 iterations and gives $T_c=0.4917$; when $\lm=1.6$, the scheme terminates
after 17 iterations and gives $T_c=0.8633$; when $\lm=4.5$, the scheme terminates after 9 iterations and gives
$T_c=2.7460$; when $\lm=6$, the scheme terminates after 8 iterations and gives $T_c=3.7137$. 

\begin{figure}[htbp]
      \centering
   \includegraphics[scale=0.8]{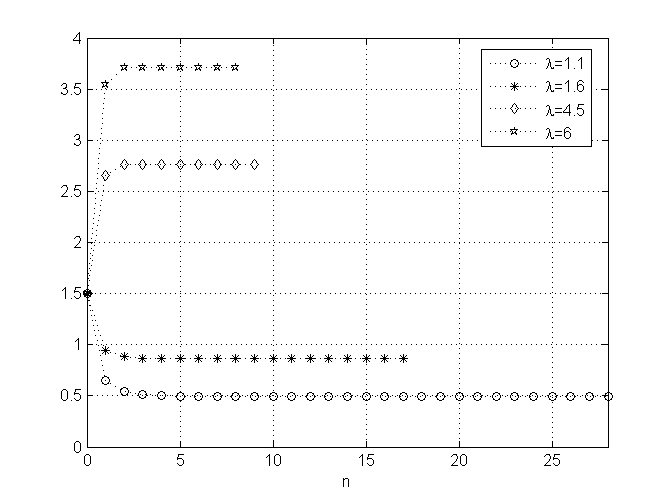}
   \caption{The behavior of the iterative sequences,  versus $\lambda$,  for the determination of the critical temperature $T_c$. All iterations start from $\tau_0=1.5$. $T_c$ increases with $\lm$.}\label{F4}
\end{figure}

\item[(ii)] Choose $\mu=1$ and $\lm=2$. Figure \ref{F5} shows the monotone dependence of $T_c$ on $\xi_m$.
 When
$\xi_m=2$, the scheme terminates after 14 iterations and gives $T_c=1.1406$; when $\xi_m=8$, the scheme terminates
after 21 iterations and gives $T_c=3.3739$; when $\xi_m=16$, the scheme terminates after 21 iterations and gives
$T_c=6.6100$; when $\xi_m=26$, the scheme terminates after 24 iterations and gives $T_c=10.6949$.

\begin{figure}[htbp]
      \centering
   \includegraphics[scale=0.8]{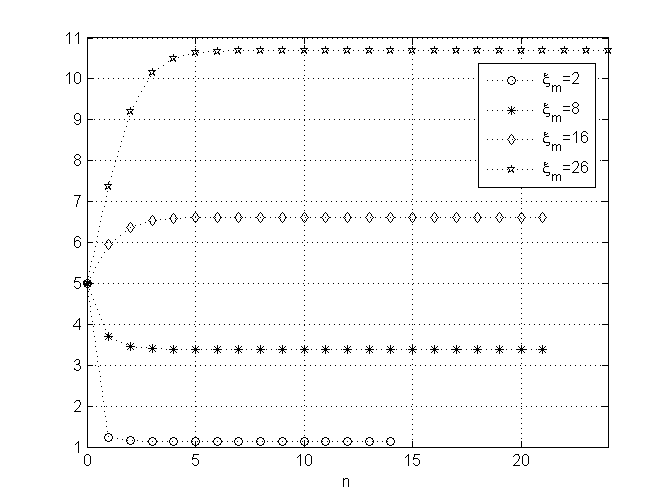}
   \caption{The behavior of the iterative sequences,  versus $\xi_m$,  for the determination of the critical temperature $T_c$. All iterations start from $\tau_0=5$. $T_c$ increases with $\xi_m$.}\label{F5}
   \label{picture-label}
\end{figure}

\item[(iii)] Fix $\xi_m=2$ and $\lm=2$. Figure \ref{F6} shows the monotone dependence of $T_c$ on $\mu$.
 When
$\mu=0.6$, the scheme terminates after 17 iterations and gives $T_c=0.9446$; when $\mu=0.9$, the scheme terminates
after 14 iterations and gives $T_c=1.0845$; when $\mu=1.2$, the scheme terminates after 11 iterations and gives
$T_c=1.2663$; when $\mu=1.5$, the scheme terminates after 10 iterations and gives $T_c=1.4875$.

\begin{figure}[htbp]
      \centering
   \includegraphics[scale=0.8]{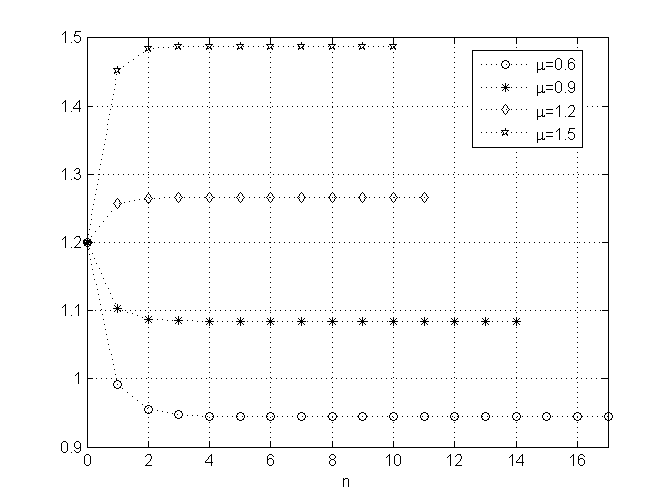}
   \caption{The behavior of the iterative sequences,  versus $\mu$,  for the determination of the critical temperature $T_c$. All iterations start from $\tau_0=1.2$. $T_c$ is enhanced by $\mu$.}\label{F6}
\end{figure}

\end{enumerate}

We next study the gap solution $\Delta$ as a function of the temperature $T$, doping $\mu$, and other parameters.
  As before a high accuracy termination 
criterion is observed in all iterative calculations.

\begin{enumerate}

\item[(i)] In Figure \ref{F7} we present plots of $\Delta$ as a function of $T$ when the set of other parameters,
$\{\lm,\mu,\xi_m\}$, is taken to be $\{2,1.5,2\}$, $\{4.5,1,2\}$, and $\{2,1,8\}$, respectively. These gap-vs-temperature
curves are typical as seen in the classical literature of BCS.

\begin{figure}[htbp]
\centering
\includegraphics[scale=0.8]{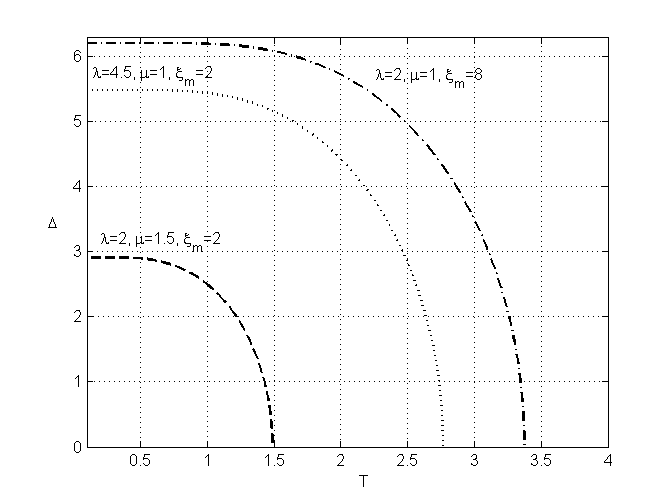}
\caption{The behavior of the gap $\Delta$ vs temperature $T$. }       \label{F7}
\end{figure}

\item[(ii)] In Figure \ref{Fig8} we present plots of $\Delta$ as a function of $\mu$ when the set of other parameters,
$\{\lm,T,\xi_m\}$, is taken to be $\{4.5,2,2\}$, $\{4.5,2.7,2\}$, and $\{2,2.7,6\}$, respectively. These curves clearly
show how doping enhances the BCS gap.  Furthermore there occurs a change-of-concavity phenomenon of $\Delta$ vs $\mu$ 
with respect to the choices of the parameters, $\lm, T, \xi_m$, unveiled by the plots.

\begin{figure}[htbp]
      \centering
   \includegraphics[scale=0.8]{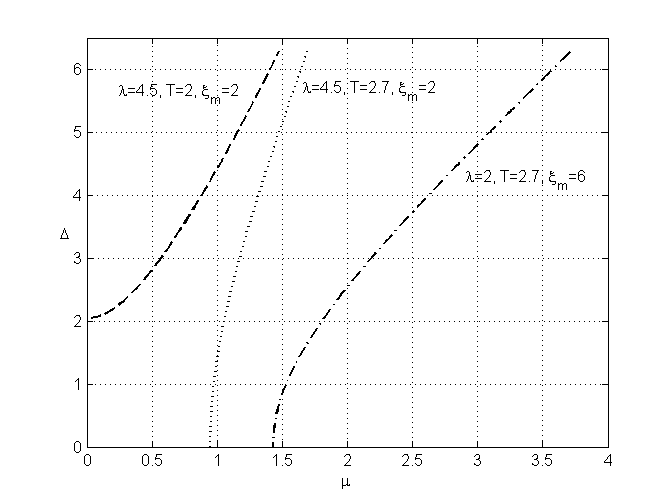}
   \caption{The behavior of the gap $\Delta$ vs doping $\mu$. }\label{Fig8}
\end{figure}

\end{enumerate}

\section{Conclusions}

In this paper we have carried out a systematic study of the doping-modified BCS gap equation in
the formalism of graphene superconductivity with regard to the existence of a positive gap solution $\Delta$ and critical
transition temperature $T_c$ in view of the changes of several physical parameters such as doping level $\mu$, pairing coupling
strength $\lm$, and the Fermi truncation energy $\xi_m$, and arrived at the following conclusions.

\begin{enumerate}

\item[(i)] Although at zero doping a positive BCS gap, $\Delta>0$,  and transition temperature, $T_c>0$, only occur when the pairing coupling is sufficiently strong,
$\lm>1$, any presence of doping, $\mu>0$, leads to the existence of a positive gap and transition temperature.

\item[(ii)] The gap $\Delta$, as a smooth function of the doping parameter $\mu\in[0,\xi_m]$, 
the pairing coupling strength $\lm>0$, and
the absolute temperature $T\geq0$, when $T<T_c$, strictly increases with
respect to $\mu$ and $\lm$, respectively, but decreases with respect to $T$, such that $\Delta\to0$ as $T\to T_c$.

\item[(iii)] The transition temperature $T_c$ as a smooth function of $\mu$ and $\lm$ strictly increases with respect to
$\mu$ and $\lm$, respectively.

\item[(iv)] The gap and transition temperature may both be computed highly effectively by globally convergent 
monotonically iterative
methods.
\end{enumerate}

In forthcoming work, we will develop and extend our methods to study more general and complicated gap equations \cite{UCN,KG,EE,UB,MN,She,CF,KHV,HZ2,GU,EMPE} arising in graphene and other superconductivity theories.

\medskip
\medskip

{\small{Xu was partially supported by National Natural Science Foundation of China under Grant No. 61201253  and
  Yang 
by National Natural Science Foundation of China under Grant No. 11471100.}}

\end{document}